\documentclass[%
,prl%
,aps%
,twocolumn%
,secnumarabic
,amssymb]{revtex4}
\usepackage{amsmath}%
\usepackage{bm}%
\usepackage{multirow}
\usepackage{ifpdf}%
\usepackage{epsf}%
\ifpdf
\usepackage[pdftex]{graphicx}
\else
\usepackage{graphicx}
\fi

\usepackage{color}

\ifpdf
\DeclareGraphicsExtensions{.pdf, .jpg}
\else
\DeclareGraphicsExtensions{.eps, .jpg}
\fi

\begin{document}
\title{Secondary Electron Emission by \\ Plasmon Induced Symmetry Breaking in  \\Highly Oriented Pyrolitic Graphite (HOPG)}

\author{ Wolfgang S.M. Werner\footnote{werner@iap.tuwien.ac.at, fax:+43-1-58801-13499}}%
\author{ Vytautas Asta\v{s}auskas}%
\author{ Philipp Ziegler}%
\affiliation{Institut f\"ur Angewandte Physik, TU Vienna, Wiedner Hauptstra{\ss}e 8-10/134, A-1040 Vienna, Austria}
\author{Alessandra Bellissimo\footnote{present address: Laboratorium f\"ur Festk\"{o}rperphysik, ETH Z\"{u}rich, Auguste-Piccard-Hof 1, Z\"urich, Switzerland}}
\author{Giovanni Stefani\footnote{present address: ISM-CNR, Via Fosso del Cavaliere 100, 00133 Roma, Italy.}}
 \affiliation{Dipartimento di Scienze, Universit\`{a} degli Studi Roma Tre,Via della Vasca Navale 84, I-00146 Rome, Italy}
\author{ Lukas Linhart}
\author{Florian Libisch}%
\affiliation{Institut f\"ur Theoretische  Physik, TU Vienna, Wiedner Hauptstra{\ss}e 8-10/136, A-1040 Vienna, Austria}

\date{\today}%
\begin{abstract}

Two-particle spectroscopy with correlated electron pairs is used to establish the causal link between the secondary electron spectrum, the $(\pi+\sigma)-$plasmon peak  and the unoccupied band structure of highly oriented pyrolitic graphite. 
The plasmon  spectrum is resolved with respect to the involved interband transitions and clearly exhibits  final state effects,  in particular due to the energy gap between the interlayer resonances along the $\Gamma$A-direction. 
The corresponding final state effects can also be identified in  the secondary electron spectrum. 
Interpretation of the results is performed on the basis of density functional theory and tight binding calculations.
Excitation of the plasmon perturbs the symmetry of the system and leads to hybridisation of the interlayer resonances with atom-like $\sigma^*$ bands along the $\Gamma A$-direction. 
These hybrid states have a high density of states as well as sufficient mobility along the graphite $c$-axis leading to the sharp $\sim$3\ eV resonance in the spectrum of emitted secondary electrons reported throughout the literature.
\end{abstract}
\maketitle 

\noindent
Van der Waals materials have recently been attracting  interest in materials science since they exhibit outstanding fundamental and technological properties and are building blocks for multilayered quasi 2D materials as well  as 3D materials and heterostructures \cite{Geim2013,Novoselov2012}.  
Graphite, being a model system for this class of materials has been most extensively studied with respect to its electronic structure, both experimentally \cite{taft,Willis1974,Skibowski1974c,Moller1982,Papagno1983,diebold1987,Maeda1988,Hoffman1990,strocov2000,Barrett2004,Barrett2004a,werner2015reflection}
and theoretically \cite{wallace,AND1958,Painter1970,Holzwarth*1982,TatarRabii1982,Marinopoulos2002a}. 

When two or more graphene layers are put on top of each other, so-called interlayer resonances form in the electronic structure which are highly dispersive along the $c$-axis and reflect the three dimensional structure of the crystal \cite{Fauster1983}. 
Distinct oscillations in the electron reflectivity are observed when measuring the reflected intensity as a function of the electron kinetic energy \cite{Hibino2008a,Feenstra2013,Jobst2015}, their number being  equal to the number of graphene layers minus one.  Interlayer states are highly transmissive for electrons coming from vacuum and have a large local density of states in between individual graphene layers. For graphite they appear as a broad band of states which strongly couple to vacuum \cite{Bellissimo2019a}. The character of such electronic  multi-quantum well states  can be qualitatively understood using the analogy to  a Fabry-P\'{e}rot interferometer in light optics \cite{Geelen2019}.  The signal employed in the above techniques, such as  elastic peak electron spectroscopy \cite{Geelen2019} and total current spectroscopy \cite{strocov2000} exclusively stems from impinging electrons which are eventually detected without having suffered any energy loss, or are absorbed in their entirety, such as in inverse photoemission spectroscopy (IPES) experiments  \cite{Fauster1983}.

Compared to the works cited above, the present paper concerns the reverse process where electrons are leaving the surface after being liberated inside the solid. This phenomenon of secondary electron emission (SEE) is of great fundamental as well as technological importance \cite{Bellissimo2019a}. 
In the past,  SEE has also extensively been employed to study the unoccupied electronic structure of graphite
\cite{Willis1974,Skibowski1974c,Moller1982,Papagno1983,Maeda1988,Hoffman1990,Hoffman1990a}. 
Obviously, for secondary electron emission, energy losses, in particular excitation and decay of plasmons \cite{Eberlein2008,shunga,Lin1997b,Papageorgiou2000,Calliari2008,Calliari2007,Guzzo2014} play an essential role. 
A striking difference between electronic structure data from SEE and the {\em elastic} techniques mentioned in the previous paragraph is that the dispersion of the interlayer resonances  is not at all observed in SEE data. Instead, a strong resonance is found in secondary electron spectra which always appears at an energy of about 3~eV above vacuum (i.e. within the energy range of the first interlayer state above vacuum). The position of this resonance shows no dispersion whatsoever  in SEE data and is found to be independent of the experimental kinematics in a substantial number of works by different authors
\cite{taft,Willis1974,Skibowski1974c,Moller1982,Papagno1983,Maeda1988,Hoffman1990,strocov2000}.

We use time-correlated two-electron spectroscopy to establish a causal
relationship between energy losses and secondary electron emission
\cite{voreadesss,scheinfeinprb47,mullultra,kruitpr44,weralcoinc} on a
sample of highly oriented pyrolitic graphite (HOPG).  In particular,
secondary electron-electron energy loss coincidence spectroscopy
(SE2ELCS) is employed (see supplemental information \cite{wersup}) to
investigate the relationship between energy losses suffered by
exciting a plasmon and the concomitant emission of a secondary
electron.  Note that for a given energy loss of the primary electron
to be feasible, corresponding initial and final states need to exist
in order to satisfy energy conservation.  The electronic transitions
taking place in resonance with the plasmon are explicitly identified
experimentally.  The experimental results highlight the influence of
the complex band structure of HOPG on the plasmon spectrum and the
ejection of a secondary electron in the course of the associated
interband transition and are interpreted with the aid of density
functional theory (DFT) and tight binding (TB) calculations
\cite{wersup}. In particular, when the symmetry of the system is
broken in our tight binding model, the resulting hybridisation of the
interlayer states with the atom-like $\sigma_2^*$-band leads to the
$\sim$3~eV resonance in the SE spectrum.

\begin{figure}[t]
{\includegraphics[width=1.1\columnwidth,trim={2cm 2.5cm 0.0cm 0.3cm},clip]{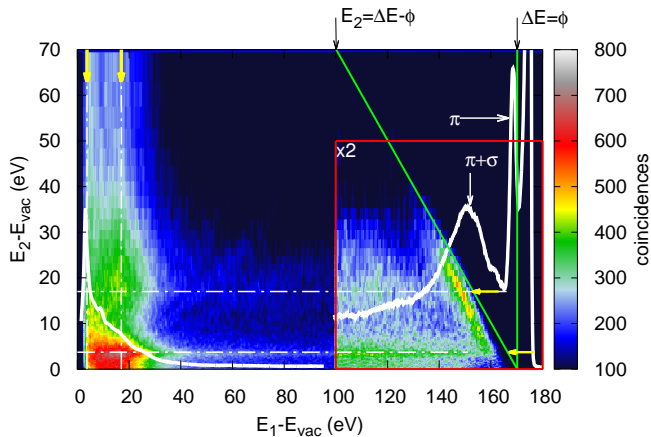}}
\caption{%
({\em color online}) 
Double differential spectrum of true coincidences for HOPG for an incident energy of 173\ eV above vacuum (energy resolution  $\delta E_1=5$\ eV).
The white curve is the singles loss-  and secondary electron spectrum. 
Indices "1" and "2" are used to indicate respectively fast and slow electrons arriving at detector 1, the hemispherical mirror analyser (HMA),  and 2, the time of flight analyser (TOF)  \cite{wersup}.
The green line labeled "$E_2=\Delta E-\phi$" indicates the minimum energy loss needed for the slow liberated electron ("2") to reach the vacuum level from the Fermi-level for a given energy loss $\Delta E$ where $\phi=4.6\ eV$ is the work function of HOPG. Here and below the yellow arrows correspond to final state energies of $E_f-E_{vac}=3.7$\ eV and 17\ eV, respectively.
}
\label{fe2dfinal}
\end{figure}
%

\begin{figure}[t]
{\includegraphics[width=1.05\columnwidth,trim={1cm 3cm 0.5cm 2cm},clip]{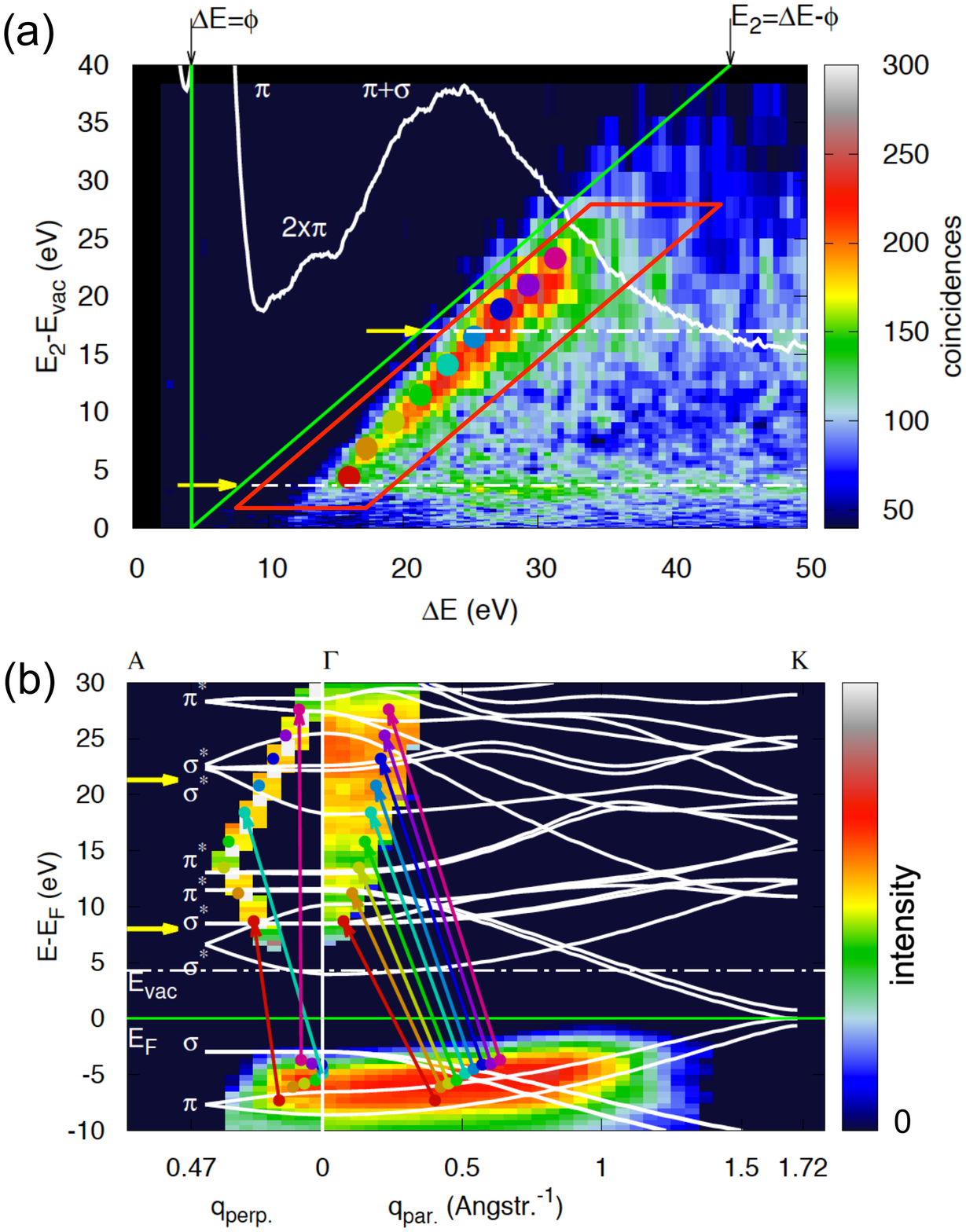}}
\caption{%
({\em color online})
(a). Coincidence spectrum of the plasmon loss in HOPG (measured with an energy resolution $\delta E_1=1.25$\ eV.)  
The white curve is the singles loss spectrum.
(b) The data in the plasmon feature in (a)  within the red parallelogram represented in $k$-space and projected on  the pertinent band structure \cite{riccardipriv} along the $\Gamma A$- and $\Gamma K$-direction (see supplemental material). The colored arrows in the panels labeled $\Gamma$A and $\Gamma$K are projections of  the initial and final states on these crystallographic directions corresponding to the colored dots in (a). 
}
\label{fl2dfinal}
\end{figure}

\begin{figure}[t]
{\includegraphics[width=0.95\columnwidth,trim={0.5cm 1.5cm 0.5cm 0.5cm},clip]{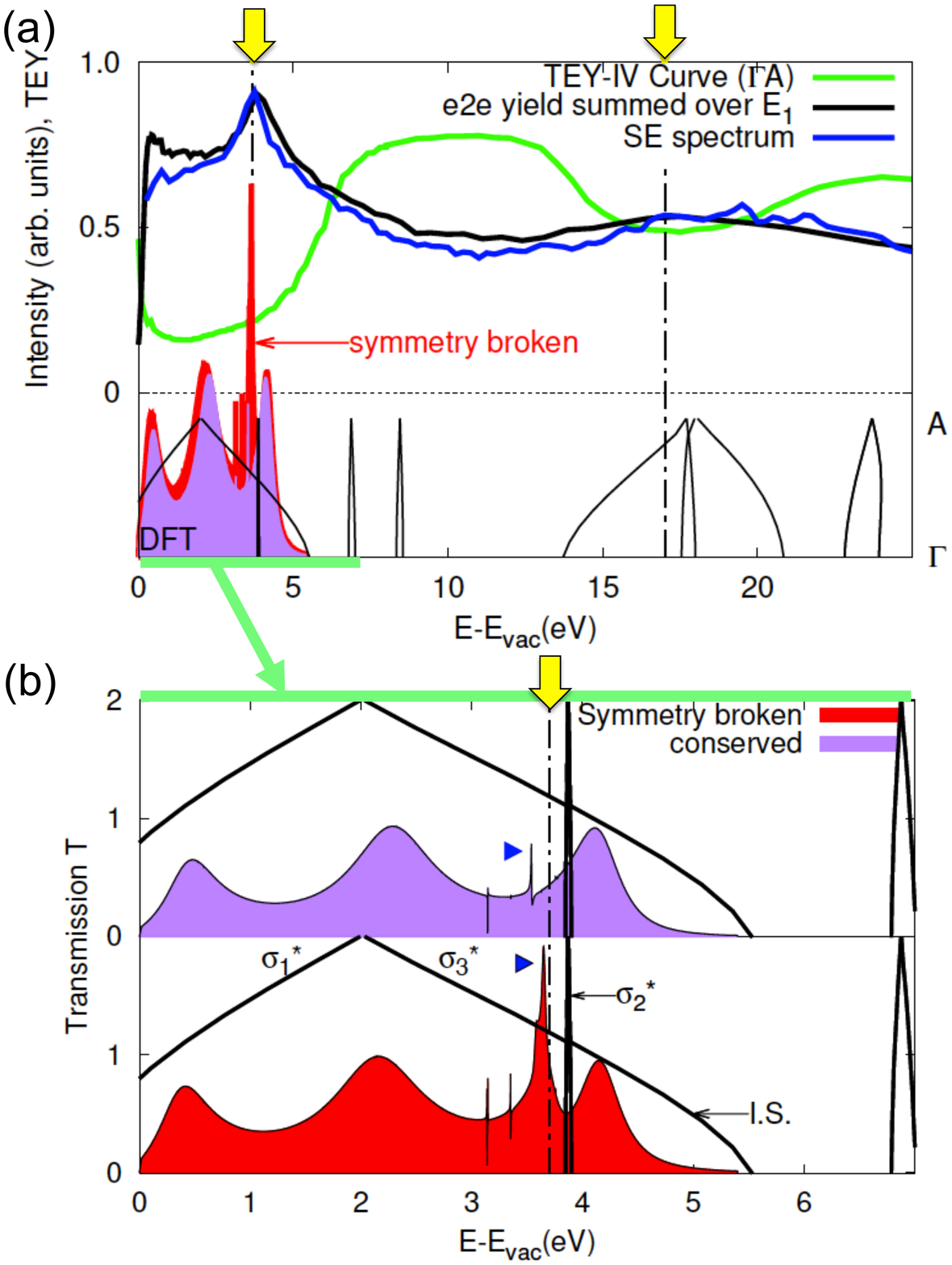}}
\caption{%
({\em color online})
(a) Green curve: total electron yield (TEY) measured in absolute units \cite{Bellissimo2019a}; Blue curve: coincidence data in Fig.\ \ref{fe2dfinal} summed over $E_1$; black curve: singles SE spectrum (white curve in  Fig.\ \ref{fe2dfinal}). The lower panel compares the band structure along  $\Gamma$A with the DFT-results for the symmetry conserved (purple) and symmetry broken (red) case.
(b) Theoretical total transmission of graphite surface
  as a function of initial energy for  symmetry-conserved (upper panel) and 
  symmetry-broken (lower panel) slab. Blue triangles mark the transmission peak caused 
  by the $\sigma_2^*$ band. In the symmetry-conserved slab, this resonance is very
  sharp, containing few electrons, while it becomes much more pronounced
  in the symmetry-broken slab due to hybridisation (see text). 
}
\label{fstripes}
\end{figure}

Our experimental results are summarized in Figs.~\ref{fe2dfinal}--\ref{fstripes}(a) showing different portions of the electron coincidence spectrum taken in  specular  reflection geometry at the Bragg maximum for a primary energy of $E_0-E_{vac}=173$~eV (see Ref\ \cite{wersup} for details). The white curve in Figs.~\ref{fe2dfinal} and \ref{fl2dfinal}(a) represents the singles electron spectrum, exhibiting the elastic peak, the $\pi-$ and $(\pi+\sigma)-$plasmon losses corresponding to minima in the real part of the dielectric function at $\hbar\omega_\pi\sim 6$~eV and  $\hbar\omega_{\pi+\sigma}\sim 23$~eV \cite{taft}. For higher energy losses plural plasmon excitation sets in. 
The secondary electron spectrum is characterized by a very sharp peak at 3.7~eV, which has been reported earlier by many authors\cite{Law1986,Fisica1986,Schafer1987,Willis1974,Moller1982,Papagno1983,Maeda1988,Hoffman1990,Hoffman1990a,strocov2000,Bellissimo2019a} and a broad shoulder at $\sim$17\ eV.

The coincidence spectrum shown in Fig.~\ref{fe2dfinal} represents the number of correlated electron pairs emitted for a given combination of energies ($E_1,E_2$). When recording the spectrum of correlated electrons in a Bragg maximum, those processes dominate in which the primary electron is first deflected along the outgoing Bragg beam followed by the inelastic process. In the deflection-loss (DL) model, one thus assumes an initial momentum of the primary electron determined by the Bragg condition {\cite{diebold1987,ruoccoprb59,Liscio2008,kheifets1998}. As a consequence, all initial and final states of the inelastic process are fixed by momentum and energy conservation \cite{wersup,ruoccoprb77}.
In other words, the coincidence experiment makes it possible to pinpoint the electronic transition of the bound electron involved in the (e,2e)-process by measuring time correlated electron pair intensities. In the present case this mainly concerns emission of a secondary electron after excitation and decay of a plasmon by the primary electron.

Three distinctly different parts can be identified in the coincidence spectrum: (1) a region of high intensity near the green line labeled $E_2=\Delta E-\phi$.
Comparison with the singles spectrum allows one to conclude that this feature corresponds to the excitation of a single plasmon, which is shown separately in Fig.\ \ref{fl2dfinal}(a); (2) horizontal stripes along the $E_1$-scale  at energies $E_2=$3.7 and 17~eV, indicated by the yellow arrows, which seem to have a counterpart along the $E_2$ scale (vertical dashed lines marked by yellow arrows). This energy region will be referred to as the plural scattering region in the following; and (3) a strong and structured peak  for energies $E_1,E_2\le20$~eV, corresponding to the cascade of secondary electrons. A distinct peak of what appears to be correlated electron emission is  seen around the point ($E_1,E_2)=(17,17)$~eV.

Fig.~\ref{fl2dfinal}(b) shows the region in phase-space corresponding to the plasmon loss in Fig.~\ref{fl2dfinal}(a), obtained by applying energy and momentum conservation 
(See Eqns. \ref{econs}-\ref{ekperp} of \cite{wersup}}) to the data within the red parallelogram in Fig.~\ref{fl2dfinal}(a). The colored arrows in Fig.~\ref{fl2dfinal}(b) indicate the interband transitions corresponding to the  energies of the fast and slow electrons marked by colored dots in  Fig.~\ref{fl2dfinal}(a).

Finally, in Fig.\ \ref{fstripes}  the total electron yield (TEY) measured in absolute units  (green curve, \cite{Bellissimo2019a}) is compared with the coincidence data in Fig.\ \ref{fe2dfinal} summed over $E_1$ (blue curve) and the singles SE spectrum (black curve). The lower panel compares the band structure along the $\Gamma$A-direction  with the DFT-results for the symmetry conserved (purple) and symmetry broken (red) case.

A striking feature of the single scattering plasmon feature is the complete absence of the $\pi$-plasmon in the coincidence data, in particular also at low energies: finite count rates appear only for $\Delta >$12 eV. This can be understood on the basis of the $k$-space representation of the data in Fig.~\ref{fl2dfinal}(b):  for the characteristic energy loss of the $\pi$-plasmon and the kinematics of the experiment, no favorable combination of initial and final states above vacuum is available which would allow a transition to take place \cite{Bellissimo2019a}. Note that our experiment mainly samples around $q_{\parallel}\sim 0.5 \rm{\AA}^{-1}$, exhibiting an energy gap between valence and conduction band.

The coincidence data also convey the fact that plural inelastic scattering proceeds via a Markov-type process \cite{werner2011moves,Bellissimo2019a}, i.e. no memory of the previous collision plays a role in subsequent processes. 
If this were not the case, i.e. if coherent plural plasmon creation took place in which the full energy loss is transferred to a single electron, intensity due to plasmon replicas should appear just below the green diagonal line at energies $(E_1,E_2)=(E_0-n\times \hbar\omega_{\pi+\sigma},n\times \hbar\omega_{\pi+\sigma}-\phi+E_b)$.  
Here $n$ is an integer and $E_b$ is the (negative) binding energy of the initial state.  
Likewise, no plasmon replicas are seen for ejected electron energies corresponding to a single plasmon loss, i.e. at energies near $(E_1,E_2)=(E_0-n\times\hbar\omega_{\pi+\sigma},\hbar\omega_{\pi+\sigma}-\phi+E_b)$. 
Instead, stripes of intensity at energies (marked by the yellow arrows) of 3.7 and 17~eV, respectively are observed. The fact that the ejected electron energy in the plural scattering region is completely independent of $\Delta E$ cannot be understood within the DL-model. 
The explanation is that when more than one inelastic process occurs, any combination of scattering angles and energy losses in individual scattering events  can lead to the net energy and momentum transfer observed for the ejected electron. 
In other words, the initial state for the process leading to ejection of the second electron is no longer determined as in the single scattering feature (see Fig.~\ref{fl2dfinal}(a)). This implies that the DL-model is only valid in the single scattering regime. 
The final state in the plural scattering regime, however, is determined by the detection geometry and energy of the slow electron. Indeed, the energies at which these stripes appear seem to  correspond to the position of the atom-like $\sigma_2^*$-bands, as well as the flat $\sigma^*$-bands 22~eV above Fermi, along the $\Gamma A$-direction (see Fig.~\ref{fstripes}). 
Note that the $\Gamma A$-direction  coincides with the symmetry axis of the TOF analyser, which measures the final state of the slow electron.

For any energy loss $\Delta E$ in the single scattering plasmon feature the probability for generating a secondary electron has a strong peak at energies $E_2$ within the plasmon feature. 
This observation highlights the fact that the final state of the scattering process corresponds to the ejection of a single bound electron. Any conceivable process in which the energy is transferred to more than one electron in the final state would lead to intensity below the single scattering plasmon feature for any reasonable energy sharing model.
Furthermore, the intensity in the single scattering feature is seen to occur for a range of slightly different binding energies (energetic distance from the green diagonal) when going from top to bottom, following the dispersion of the initial state. This is indeed confirmed by the representation of the single scattering data in $k$-space, Fig.~\ref{fl2dfinal}(b). 
A faint minimum is seen in the plasmon loss feature  in Fig.~(\ref{fe2dfinal}) near $(E_1,E_2)=(154,10)$~eV, corresponding to the energy gap in between the strongly dispersing interlayer bands along $\Gamma A$ indicated by the yellow arrows in Fig.~\ref{fl2dfinal}(b). 
Indeed, a minimum in the final state intensity is also observed in the plasmon spectrum in Fig.~\ref{fl2dfinal}(b) at $E-E_F\sim$15\ eV.

These findings show that for a material with a complex band structure, such as graphite, the excitation of a plasmon and the associated interband transitions are both essential parts of the same coherent process, a plasmon-assisted interband transition \cite{kouzakov2012} leading to ejection of the bound electron into an excited state. This picture supports the momentum-exciton model for the plasmon \cite{Ferrell1958,Bellissimo2019a} as a coherent excitation of a (rather small) number of electron hole pairs behaving as a quasi-particle with a well-defined energy and momentum. 
Note that the typical number of electrons participating in a plasmon  is about five \cite{Ferrell1958,egri} and that the range of possible energies above vacuum occupied by the ejected electron is limited by the plasmon energy. This implies that the width of the secondary electron peak is essentially governed by the density of the solid state electrons.

Finally, the energies of the secondary electrons emitted along the
graphite $c$-axis (stripes in the plural scattering region marked with
yellow arrows) and the sharp peak in the SE spectrum at 3.7~eV
exhibiting a surprising lack of dispersion
\cite{Law1986,Fisica1986,Schafer1987,Willis1974,Moller1982,Papagno1983,Maeda1988,Hoffman1990,Hoffman1990a,strocov2000,Bellissimo2019a}
deserve to be discussed. These features appear energetically very
close to the flat $\sigma_2^*$- bands within the interlayer resonances
along $\Gamma A$ (Fig.\ref{fl2dfinal}(b)). To understand their origin
we use DFT calculations for bulk graphite as well as a surface slab to
parameterize a tight binding model \cite{Linhart2019}. We can then
simulate the transmission of the secondary electron from a Bloch
state inside the solid, i.e. the final state of the inelastic
scattering process with the incoming electron to a free vacuum state
(as given by Eq.~(\ref{eq:Tscatt}) in the supplemental material).
Given an unperturbed graphite slab, we find, as expected by the flat
nature of the $\sigma_2^*$ bands, a very sharp resonance (see blue
triangle in upper panel of Fig.~\ref{fstripes}(b)) that does not
contribute significantly to the overall signal, as its area is
vanishingly small. We next aim to model the symmetry breaking induced
by the incidence of the primary electron and the emerging plasmon in
the simplest way possible: we induce a symmetry-breaking at
the surface of the slab by a local potential $V_{\delta \mathrm{sp^2}}
\approx 0.3 eV$ added to a single of the three equivalent
$\mathrm{sp}^2$ orbitals of each carbon atom of the surface
layer. Such a term breaks the $D_{6h}$ symmetry of graphite,
essentially locally eliminating the three-fold symmetry.  The induced
hybridisation between the $\sigma_2^*$ bands and
$\sigma_{1,3}^*$-interlayer bands substantially enhances and broadens
the resonance. The resulting hybrid state exhibits the high density of
states of the flat band, implying that it is a favorable state for an
initially bound electron to reach a final state above the vacuum
level. The hybrid state also has the high mobility of the interlayer
state which efficiently couples to vacuum, allowing it to escape from
the surface. We have verified numerically that a similar symmetry
breaking in bulk HOPG leads to transmission from $\sigma_2^*$ to
$\sigma_{1,3}^*$ modes (and vice versa) in bulk transport
perpendicular to the layers.

Since the parametrization of our tight-binding model only extends to
about 18~eV, the origin of the second stripe in the coincidence data
at $E-E_F\sim$17~eV could not be verified. It is conjectured however,
that a similar hybridisation mechanism plays a role in that case as
for the lower lying interlayer band around 3.7~eV. The peak of
apparent correlated emission at $(E_1,E_2)=(17,17)$~eV then is a
direct consequence of the hybridisation in that also the unoccupied
flat band at around 17~eV (above vacuum) is a favorable final state
and by virtue of the high mobility of the hybridized state also leads
to a strong peak in the secondary electron cascade. For the primary
energy $E_0=$173~eV employed here, plural inelastic processes can lead
to (incoherent) creation of several secondary electrons.  A part of
these SEs may actually escape and give rise to multi-electron
detection events with prefered energies of 17~eV.  As discussed above,
this proceeds via a Markov sequence of events leading to a strong
preference of final states with these energies in the course of
(incoherent) plural inelastic scattering and generation of secondary
electrons. Therefore, while the peak at $(E_1,E_2)=(17,17)$~eV appears
to be due to correlated electron emission, it rather is an incoherent
increase in the electron pair intensity due to the aforementioned
process.  A similar peak is also expected at $(E_1,E_2)=(3.7,3.7)$ but
could not be observed since the lowest energy along the $E_1$-scale
that can be reached in the coincidence experiment is
$\sim$10~eV.

In summary, the $(\pi+\sigma)$-plasmon in graphite has been resolved
with respect to the involved electronic transitions.  Formation of a
hybrid state as a consequence of plasmon induced symmetry breaking
provides an explanation for the strong resonance observed in
SE-spectra in the literature (at 3.7~eV in the present work).  Most
importantly, it explains the difference in band structure measurements
using elastic processes and techniques which involve creation of
plasmon.  To fully appreciate this point the reader is referred to
Fig. 4 in Ref.~\cite{Maeda1988} which shows a direct comparison of
IPES and SEE data, the dispersion of the interlayer state completely
lacking in the latter.  The results therefore indicate that the
inverse-LEED (Low Energy Electron Diffraction) formalism which is
often employed to interpret secondary electron spectra
\cite{luethbook} should be complemented to account for many body
processes, since the phenomenon of secondary electron emission cannot
be fully described on the basis of the one-electron band structure.

%


{\bf Acknowledgments}\\ The authors express their gratitude to
P. Riccardi and E. Krasovskii  for fruitful discussions and 
P. Riccardi for providing detailed band structure data of graphite.
Financial support by the FP7 People: Marie-Curie Actions Initial
Training Network (ITN) SIMDALEE2 (Grant No. PITN 606988) is gratefully
acknowledged.  FL and LL acknowledge support by FWF program I
3827-N36. The computational results presented have been achieved using
the Vienna Scientific Cluster (VSC).

\bibliography{/Users/werner/ownCloud/tex/Allrefnewest,mycollection}
\newpage .
%
\begin{center}
{\Large Supplemental Material for {\bf "Secondary Electron Emission by \\ Plasmon Induced Symmetry Breaking in  \\Highly Oriented Pyrolitic Graphite (HOPG)"}}\\
{\it Wolfgang S.M. Werner}, {\it Vytautas Asta\v{s}auskas},\\ {\it Philipp Ziegler}, %
{\it Alessandra Bellissimo,}\\ {\it Giovanni Stefani}, {\it Lukas Linhart}\\ and {\it Florian Libisch}.%
\end{center}


\section{Experimental}
The sample was a HOPG crystal of ZYB quality with a nominal mosaic spread of $(0.8\pm 0.2)^\circ$, mechanically exfoliated in air and  annealed for several hours at 500$^\circ$C  in the UHV system with a base pressure of 2$\times 10^{-10}$\ mbar housing the  spectrometer.  During the measurement, the pressure rises to 5$\times 10^{-10}$\ mbar.

\begin{figure}[t]
{\includegraphics[width=0.8\columnwidth]{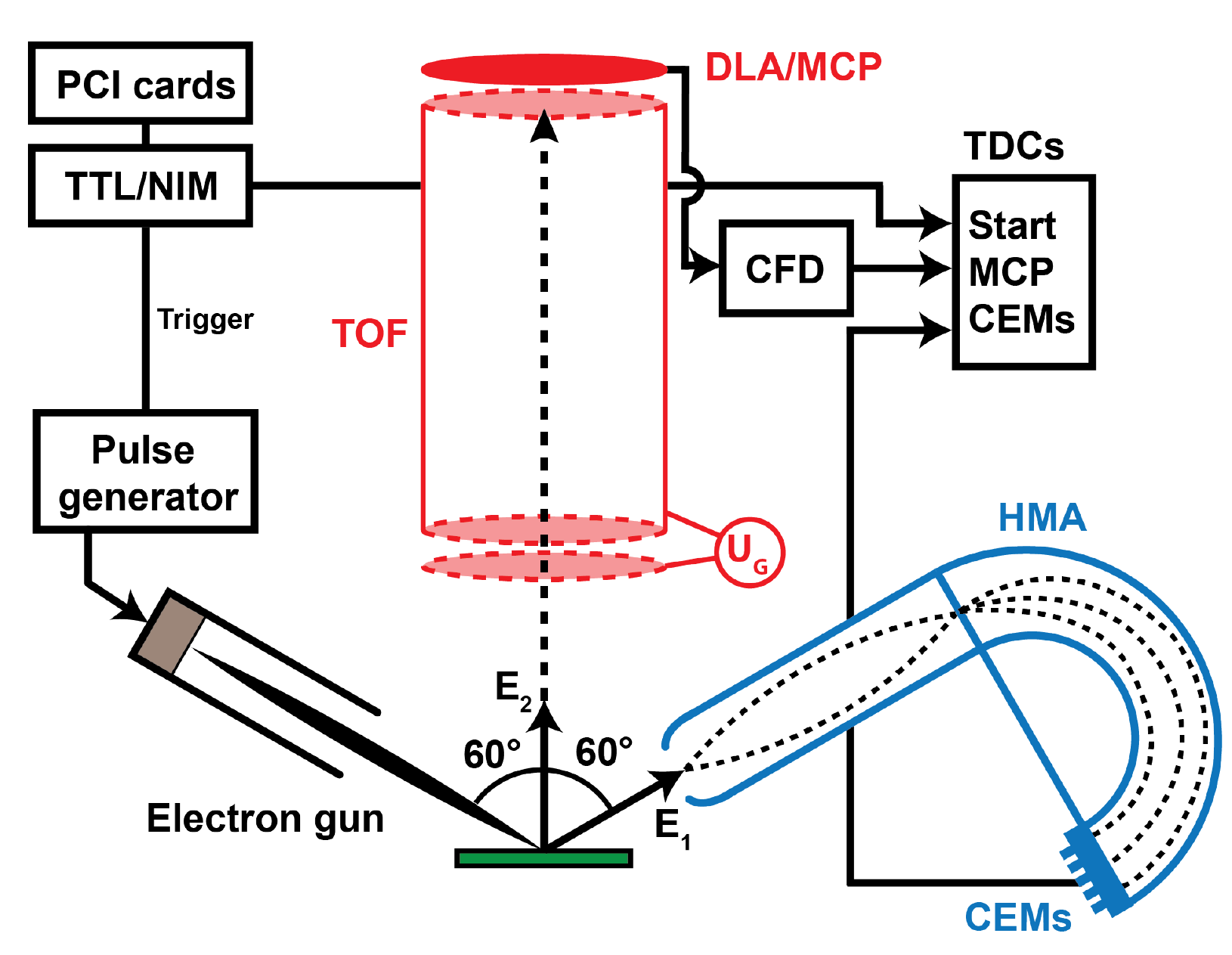}}
\caption{%
({\em color online}) 
Schematic illustration of the experimental setup. (see text).
}
\label{fexp}
\end{figure}

The employed coincidence spectrometer is schematically shown in  Fig.\ \ref{fexp}. Arrival-time differences of electron pairs are recorded for a hemispherical mirror analyser (HMA, measuring the fast scattered electron, subscript '1'),  and a time of flight analyser (TOF, measuring the slow ejected electron, subscript '2').  
While the HMA has a constant energy resolution $\delta E_1$ (equal to 5\ eV and 1.25\ eV for the data presented in Fig.~\ref{fe2dfinal} and Fig.~\ref{fl2dfinal} respectively), the energy resolution of the TOF-analyser is about $\delta E_2=1$~eV at an energy of 20~eV, while it is several tens of an eV at  an energy of 200\ eV. 
The white curve in Fig.~\ref{fe2dfinal} representing the singles energy loss and SE spectrum was measured  with the HMA with a pass energy of 5\ eV corresponding to an energy resolution of $\delta E_1=0.25$~eV. 
The lowest energy which can be recorded by the HMA depends on the employed pass energy and is equal to 10\ eV for the coincidence data in  Fig.~\ref{fe2dfinal}, while it is 
0.5\ eV for the singles spectrum (white curve).
Coincidence spectra were measured at a primary energy $E_0-E_{vac}=173$~eV, corresponding to a Bragg maximum for this geometry, after carefully aligning the sample in the specular reflection geometry, by directing the Bragg beam into the central angle of the HMA. 

The coincidence spectra are obtained with a continuous electron beam in order to increase the effective coincidence rate. Flight times for each energy measured by the 
hemispherical analyser are measured before the coincidence measurement using a pulsed electron beam. In this way, the flight time of an electron in a correlated pair 
reaching the time of flight detector during the coincidence measurement is calibrated. Coincident flight time spectra for each energy measured with the HMA are 
superimposed on a flat background of false coincidences which is subtracted, leading to the spectra of true coincidences as shown in  Fig.~\ref{fe2dfinal} and 
Fig.~\ref{fl2dfinal}. The current incident on the sample during the coincidence measurement amounted to 70~fA, obtained by comparing elastic peak intensities with picoamperemeter results down to the nA range. In this way, the spectrometer transmission and other experimental factors are calibrated giving absolute values at any lower current using the elastic peak intensity.
The acquisition time for the coincidence data shown in Fig.~\ref{fe2dfinal} as well as for the coincidence data in Fig.~\ref{fl2dfinal} amounted to approximately 14 days.

 Note that both analysers are located in the scattering plane. The opening angles of the analysers amount to $\Delta\Omega_{HMA}=10^\circ$ and $\Delta\Omega_{TOF}=12^\circ$. Owing to the employed experimental geometry (see Fig.~\ref{fexp}), the axis of the TOF analyser is aligned with the $\Gamma A$ crystallographic direction (i.e. the $c$-axis in graphite), while the HMA intensity consists of components along the $\Gamma A$ as well as the $\Gamma K$ and $\Gamma M$-crystallographic direction. 
 Due to the wide opening angle of the hemispherical analyser the momenta sampled in the occupied band structure extend to about 1$\rm{\AA}^{-1}$.  Since the $\Gamma M$ and $\Gamma K$ band structure are similar to each other near the $\Gamma $-point and cannot be distinguished within our experimental energy and angular resolution we only present and discuss the $\Gamma K$-portion in the text for clarity. While for the occupied states, the $q_\parallel,q_\perp$-region sampled in phase space is mainly governed by the angular acceptance of the HMA, the acceptance of the TOF governs the portion of final state phase space, which becomes particularly narrow for the $\Gamma A$-direction owing to the increase of $k_\perp$ inside the solid (see below). 


   
\section{Data interpretation}
In the specular reflection geometry,  diffraction of the incoming beam by  the graphene planes gives rise to a Bragg beam oriented exactly along the specular direction inside the solid, which then can induce an inelastic process. In the UV-energy range an energy loss is predominantly accompanied by an electronic transition,  which leads to liberation of a second electron inside the solid. For sufficiently large losses, this electron may overcome the surface potential barrier and is ejected from the surface.  This deflection-loss (DL) model   \cite{diebold1987,ruoccoprb59,Liscio2008,kheifets1998} implies that the scattering kinematics of the inelastic collision are fully determined.
In other words, under such conditions, the final state of the correlated electron pair  in an (e,2e) experiment is fixed by the energy and  direction of detection and together with the known  state of the primary electron, the initial state of the ejected solid state electron  is fully determined. 
Here we designate the scattered and ejected electron by the indices "s" and "e", while in the main text we merely distinguish between events where electrons are detected in 
analyser 1 and 2  and label the energy scales accordingly. 
This is strictly speaking correct and necessary because of the indistinguishability of electrons but generally identifying electron 1 with the scattered (fast) electron and electron 2 with the ejected (slow) electron will be correct in many cases.
The binding energy $E_b(\vec{q})$ of the bound electron is found by requiring that the energy loss of the primary electron  $\Delta E=E_0-E_s$ --where the index "0" indicates the primary electron-- is used to liberate the bound electron from the solid, by overcoming the work function $\Phi$, and that it is ultimately ejected from the solid with an energy $E_e$:
\begin{equation}
\label{econs}
\Delta E=E_0-E_s=E_e+\Phi-E_b(\vec{q}),
\end{equation}
where the  binding energy is counted from the Fermi level and is negative. Momentum conservation yields for the momentum of the bound electron:
\begin{equation}
\label{emomcons}
\vec{q}=\vec{k}_e-(\vec{k}_0-\vec{k}_s)+\vec{G}=\vec{k}_e-\Delta\vec{K}+\vec{G},
\end{equation}
where $\vec{G}$ is a reciprocal lattice vector and $\Delta \vec{K}$ is the momentum transfer. The above equation holds inside the solid. 
To convert the momenta inside the solid to the momenta measured in vacuum (or vice versa),  the increase of the energy by the inner potential $U_i$ when the electron crosses the surface potential barrier needs to be accounted for. Using atomic units ($\hbar=m_e=e=1$) the relationship between the perpendicular  component of momentum inside and outside the solid can be expressed as follows:
\begin{equation}
\label{ekperp}
\frac{k^2_{\perp,in}}{2}=\frac{k^2_{\perp,out}}{2}+U_i,
\end{equation}
while the parallel momentum is conserved as the electron penetrates the barrier.

The inner potential $U_i$ has been determined by measuring the elastic peak intensity as a function of the vacuum energy $E_v$ and identifying the $n$-th order Bragg maximum $E_{Bragg}(n)$. Using the Equation \cite{ruoccoprb59}:
\begin{equation}
E_{Bragg}(n)=\Big\{\frac{n^2\pi^2}{2d^2}-U_i \Big\} \frac{1}{\cos^2\alpha},
\end{equation}
which is written here in atomic units with $d$  the interlayer distance and $\alpha$ the vacuum polar angle of incidence and emission, we have plotted the Bragg energies in vacuum against $n^2$ and determined the inner potential from the axis offset after fitting the data to a straight line. The value of $U_i=16.1~eV$ obtained in this way is in good agreement with values found in the literature (see Ref.~\cite{ruoccoprb59} and references therein).

Using Eqns. \ref{econs}--\ref{ekperp}, values of the initial and final state energies and momenta are determined for each pixel in the data of Fig.~\ref{fl2dfinal}(a). The corresponding intensity in each pixel is added to a histogram in $k$-space, eventually leading to the results shown in Fig.~\ref{fl2dfinal}(b). The portion of phase space along the $\Gamma$A-direction becomes very narrow due to the significant increase of the perpendicular momentum component when the electron feels the inner potential of $U_i=16.1$~eV.

\section{Theory}

Given the complexity of the inelastic scattering processes inside
HOPG, we do not model this process beyond the energy and momentum
conservation considerations outlined above. Instead, we aim to
elucidate the ejection process of the ejected electron by considering
its scattering from an initial state inside the HOPG target (the final
state of the inelastic scattering process) to a final propagating
state outside of the solid (resulting in a signal at the detector). We
first model bulk graphite and a graphite surface slab using density
functional theory (DFT) employing the VASP software package
\cite{Kresse1996}.  We then use a Wannier localization
procedure\cite{Marzari1997,Souza2001} to obtain local tight-binding
parameters for the bulk and the surface slab following
\cite{Linhart2019}. We verify that our tight binding model reproduces
the full DFT single-particle band structure. Since we are
interested in electron emission at higher energies, we take great care
in obtaining a converged band structure at energies up to 20 eV in
VASP using appropriate convergence thresholds. Given the strong
entanglement of higher-lying virtual bands, our tight binding model
only correctly reproduces bands up to 18 eV.

Using our Wannier representation, we compose a tight-binding
Hamiltonian $H$ containing a large number of bulk layers, four surface
layers from the slab and free electron states outside the slab. We
obtain a scattering problem by introducing open boundary conditions at
both sides by infinite waveguides representing graphite Bloch
states (incoming lead) and free particle states (outgoing lead).  The
Green's function of this system is given by
\begin{equation}
  G(E) = \left[H - E - \Sigma_{Gr}(E) - \Sigma_V(E)\right]^{-1},
\end{equation}
where the self-energies $\Sigma_{Gr}$ ($\Sigma_V$) represent the open
boundary conditions of bulk graphite (vacuum).  We solve the
scattering problem using our modular recursive Green's function
approach \cite{Libisch2012} to determine the probability $T(\mathbf
k_B)$ of an electron in an initial Bloch state $\mathbf k_B$ with
energy $\varepsilon_{\mathbf k_B}$ inside the solid transfering to a
propagating vacuum state,
\begin{equation}\label{eq:Tscatt}
  T(\mathbf k_B) =\int_D d\mathbf k_v \left|\left< \phi_{\mathbf k_B}\right| G(\varepsilon_{\mathbf k_B}) \left|\mathbf k_v\right> \right|^2\delta(\varepsilon_{\mathbf k_B}-\mathbf k_v^2/2m )
\end{equation}

where the integral over the vacuum wave vector $\mathbf k_v$ goes over those final states that can reach the detector in the
experimental setup, and the delta function ensures energy conservation.

\end{document}